\documentclass[twocolumn,superscriptaddress,pra,aps,showpacs]{revtex4-1}
\usepackage{amsmath,mathrsfs,amsbsy,color,graphicx,bm,amsthm,amsfonts}
\usepackage{newlfont}
\usepackage{times}
\usepackage{graphicx}
\usepackage{ulem}
\usepackage{braket}

\newcommand{\be}{\begin{equation}}
\newcommand{\ee}{\end{equation}}
\newcommand{\br}{\begin{eqnarray}}
\newcommand{\er}{\end{eqnarray}}
\newcommand{\tr}{\text{Tr}}

\begin{document}
\title{Strong Subadditivity Lower Bound and Quantum Channels}
\author{L. R. S. Mendes }
\affiliation{Instituto de F\'isica de S\~ao Carlos, Universidade de S\~ao Paulo, 13560-970, S\~ao Carlos, SP, Brazil}
\affiliation{Instituto de F\'isica Gleb Wataghin, Universidade Estadual de Campinas, 13083-859, Campinas, SP, Brazil}
\author{M. C. de Oliveira}
\email{marcos@ifi.unicamp.br}
\affiliation{Instituto de F\'isica Gleb Wataghin, Universidade Estadual de Campinas, 13083-859, Campinas, SP, Brazil}

\begin{abstract}
We derive the strong subadditivity of the von Neumann entropy with a strict lower bound dependent on the distribution of quantum correlation in the system. We investigate the structure of states saturating the bounded subadditivity and explore its consequences for the quantum data processing inequality. The quantum data processing achieves a lower bound associated with the locally inaccessible information.
\end{abstract}

\maketitle

\section{Introduction}
In information theory the central constraint on how information can be distributed among parties is given by the subadditivity inequalities \cite{TC,mwilde}. Given two random variables $X:\{x\}$ and $Y:\{y\}$, assuming the values $x$ and $y$ with probabilities $p_x$ and $p_y$, respectively, the weak subadditivity is given by
\begin{equation}
    H(X,Y)\leq H(X)+H(Y),
\end{equation}
in terms of the Shannon entropies $H(X)=-\sum_{X:\{x\}} p_x \ln p_x$,
$H(Y)=-\sum_{Y:\{y\}} p_y \ln p_y$  and the joint entropy $H(X,Y)=-\sum_{X:\{x\},Y:\{y\}} p_{x,y} \ln p_{x,y}$. This inequality,
in essence, states the positivity of the mutual information 
$
I(X:Y)\equiv H(X)+H(Y)-H(X,Y)\geq 0,
$
which bounds the correlations between $X$ and $Y$.
On the other hand, the strong subadditivity
\begin{equation}
    H(X,Y,Z)+ H(Y)\geq H(X,Y)+H(Y,Z),
\end{equation}
imposes the positivity of the conditional mutual information defined by
$ I(X:Z|Y)\equiv H(X,Y)+H(Y,Z)-H(X,Y,Z)- H(Y)\geq 0$.
As simple as they look these two positivity bounds govern what one can or cannot do in communication since they are related to the communication channel capacity and essentially all other relevant inequalities in information theory\cite{TC}.

Quantum information theory is concerned with information processing and  tasks performing in the quantum regime, and the main quantity for that is the von Neumann entropy
\be
S(\rho) = -\tr\left(\rho \log \rho\right). 
\ee 
Similarly to classical information theory, the subadditivity for the von Neumann entropy is enormously relevant. The weak subadditivity states as
\be
S(A,B)\leq S(A)+S(B),\label{WSA}
\ee
being  $S(A)\equiv S(\rho_{A})$, and $S(A,B)\equiv S(\rho_{AB})$. We may also define a mutual information as $I(A:B)=S(A)+S(B)-S(A,B)$, being always positive. Moreover the strong subbaditivity (SSA), proved by Lieb and Ruskai \cite{lib} gives that
\be
S(A,B,C) + S(B) \leq S(A,B) + S(B,C),\label{SSA}
\ee
being $S(ABC)\equiv S(\rho_{ABC})$. This inequality holds for a tripartite system with density matrix $\rho_{ABC}$ living in ${\cal{H}}_{ABC}={\cal{H}}_A\otimes {\cal{H}}_B\otimes {\cal{H}}_C$, and each of the reduced density matrices were taken by a partial trace of the density matrix of the role system in a way that $\tr_{BC}(\rho_{ABC}) =  \rho_{A}$. Similarly to the classical instance the SSA of the von Neumann entropy reduces to the positivity of the conditional mutual information in the form
\begin{equation}
I(A:C|B)=S(A,B)+S(B,C)-S(B)-S(A,B,C)\geq 0.\label{ruskLieb}
\end{equation}
Its relevance is so extensive that it is applied even outside communication scenarios e.g., to the search of lower bounds for the free energy in a many body problem context \cite{poulin}.

But more specifically, there are some implicit problems in the distinction of genuinely quantum from classical correlation when the mutual information for quantum systems is given by (\ref{WSA}). This is because the mutual information can be more precisely defined as
\begin{equation}
I(A:B)=S(A)-S(A|B),\label{mutual}
\end{equation}
where $S(A|B)$ is the conditional information on $\rho_A$ given the knowledge of state $\rho_B$. The form $S(A|B)=S(A,B)-S(B)$ is borrowed from classical information theory, where it is strictly valid. In quantum information to have some knowledge about the quantum state, some measurement must be performed and therefore a more appropriate form of conditional entropy must be employed \cite{zurek}. It turns out that when such an approach is considered some intriguing relations between entanglement and quantum correlation in general emerge \cite{winternew}, leading to  a weak monotonicity relation of the von Neumann entropy \cite{Fan11} with additional restrictions on the balance of quantum correlations in the system as measured by the entanglement of formation (Eof)\cite{eof} and the quantum discord (QD) \cite{zurek}. Since the standard weak monotonicity is equivalent to the SSA, it would be interesting to extend the discussion to understand the existence of possible new bounds.

In this work we further investigate this problem by deriving a strong subadditivity relation from the bounded weak monotonicity. We show that it intrinsically involves new bounds which allows distinction of genuine quantum correlations. In \cite{hayden} it was shown that the structure of states that saturates the SSA (\ref{SSA}) is given by
\begin{equation}
\rho_{ABC} = \bigoplus_{j} q_{j} \rho_{Ab^{L}_{j}}\otimes \rho_{b^{R}_{j}C},
\end{equation}
where $\rho_{Ab^{L}_{j}} \in {\cal{H}}_{A}\otimes {\cal{H}}_{b^{L}_{j}}$ and $\rho_{b^{R}_{j}C} \in {\cal{H}}_{b^{R}_{j}}\otimes {\cal{H}}_{C}$, with probability distribution $\{q_{j}\}$, such that the Hilbert space be decomposable as ${\cal{H}}_{B}=\bigoplus_{j} {\cal{H}}_{b^{L}_{j}}\otimes {\cal{H}}_{b^{R}_{j}}$. This structure was enormously relevant for the derivation of a hierarchy of independent inequalities for the von Neumann entropy \cite{cadney}. Here we extend this analysis to understand the structure of states saturating the bounded SSA. Moreover we apply the derived bound to the quantum data processing inequality. The paper is organized as follows. In Sec. II we develop the new bounded SSA and discuss its implications in several instances. In Sec. III we investigate the structure of states saturating the bounded-SSA. Finally in Sec. IV we apply the bounded-SSA to understand the imposed restrictions on the quantum data processing inequality \cite{schuma}.
In Sec. V a conclusion encloses the paper.

\section{Bounded strong subadditivity}
The mutual information (MI) in  terms of the von Neumann entropy measures the amount of information shared by two quantum systems $A$ and $B$. In other words it quantifies the total amount of correlations (quantum and classical) of a bipartite quantum state $\rho_{AB}$. It is given by eq. (\ref{mutual}).  Assuming the extension of the classical form for the conditional entropy to the quantum case as being $S(A|B)=S(A,B)-S(B)$, we end up with
\begin{equation}
I(A:B)= S(A) + S(B) - S(A,B).\label{Iq1}
\end{equation} 
In  contrast, by taking into account the fact that in quantum systems for the description of the conditional entropy $S(A|B)$ the prior knowledge about $\rho_B$ is achieved by some kind of measurement one obtains
\begin{equation}
J^{\leftarrow}_{A|B} =\max_{\left\{ \Pi
_{k}\right\} }\left[S(A)-\sum_{k}p_{k}S({A|k})\right],\label{2}
\end{equation}
where $S({A|k})\equiv S(\rho_{A|k})$ is the conditional entropy after a measurement in $B$, where $\rho_{A|k}=\mathrm{Tr}_{B}(\Pi _{k}\rho_{AB} \Pi _{k})/\mathrm{Tr}_{AB}(\Pi_{k}\rho_{AB}\Pi_{k})$ is the reduced state of $A$
after obtaining the outcome $k$ in $B$. $\{\Pi_k\}$ is a complete set of positive operator valued measurement resulting in the outcome $k$ with probability $p_k=\mathrm{Tr}_{AB}(\Pi_{k}\rho_{AB}\Pi_{k})$. In this case, since a measurement might give different results depending on the basis choice, a maximization is required.
Thus $J_{A|B}$ measures the amount of mutual information accessible by local measurement in $B$ only.
Due to that distinction in definition one can quantify the amount of information not accessible by local measurements in $B$ by
\begin{equation}
\delta^{\leftarrow}_{AB} = I({A:B}) - J_{A|B}^{\leftarrow},\label{qdisc}\end{equation} the so-called quantum discord \cite{zurek}.

For an arbitrarily mixed tripartite system state $\rho_{ABC}$ there exists an important relation  known as
the Koashi-Winter inequality \cite{winternew,Fan11} and given by
\be
E_{AB} \leq \delta^{\leftarrow}_{AC} + S_{A|C},
\ee 
where $E_{AB}$ quantifies the entanglement of formation (Eof) between $A$ and $B$ and $\delta^{\leftarrow}_{AC}$ is the quantum discord (QD) between $A$ and $C$ (given measurements in $C$).
In fact it is possible to show \cite{Fan11} that in general 
\be\label{eq:3}
S(B) + S(C) + \Delta \leq S({A,B}) + S({A,C}),
\ee
where $\Delta$ is the balance of correlations in a tripartite system, and is given by  
\be
\Delta=E_{AB}+E_{AC}-\delta^{\leftarrow}_{AB} - \delta^{\leftarrow}_{AC}.
\ee 
Pure quantum states $\rho_{ABC}$ necessarily satisfy $S(B)=S({A,C})$ and $S(C)=S({A,B})$ and saturate (\ref{eq:3}) in a way that $\Delta=0$, or
\be \label{eq:4}
E_{AB}+E_{AC}=\delta^{\leftarrow}_{AB} + \delta^{\leftarrow}_{AC}.
\ee
The balance of quantum correlations above can be viewed as a conservation relation that states that the entanglement of formation of a bipartite system is going to be increased or decreased by the same amount that the quantum discord of the same bipartite system in relation to a part of the pure tripartite global state.

By adding an ancilla $R$ so that the global state of the system $\rho_{ABCR}$ is purified the system partitions entropies relate as $S(R) = S(A,B,C)$ and $S(A,R) = S(B,C)$. To write $\Delta$ for the extended system we use the conservation relation (\ref{eq:4}) between the Eof and the QD for the two subsystem $\mathcal{H}_{AR} = \mathcal{H}_{A}\otimes \mathcal{H}_{R}$ and $\mathcal{H}_{ABC} = \mathcal{H}_{A}\otimes \mathcal{H}_{BC}$ so that
\be\label{eq:5}
S(A,B,C) + S(B) + \widetilde{\Delta} \leq S(A,B) + S(B,C) ,
\ee
where  
\be
\widetilde{\Delta} = E_{AB} - E_{A(BC)} + \delta^{\leftarrow}_{A(BC)} - \delta^{\leftarrow}_{AB}.
\ee 
The inequality in Eq. (\ref{eq:5}) is similar to the SSA, but for the additional term $\widetilde{\Delta}$. Since $\widetilde{\Delta}$ can take both positive and negative values it can be a stronger or weaker bound to the SSA, and therefore we call it as the bounded SSA, or b-SSA for short. Since the positivity of the conditional mutual information $I(A:C|B)$ (\ref{ruskLieb}) was independently proved by Lieb and Ruskai \cite{lib}, in fact in (\ref{eq:5}) effectively one has to take it in (\ref{eq:5}) as
\begin{equation}
I(A:C|B)\geq \max\{0,\widetilde{\Delta}\}.\label{ruskLiebnew}
\end{equation}





In transitioning from (\ref{eq:3}) to (\ref{eq:5}) there is a change in signs in the balance of quantum correlations. So that the difference of the positivity of the $\Delta$  to $\widetilde{\Delta}$ does not depend only on the difference between the QD and the Eof of the same bipartitions, becoming more complex to evaluate in general. Specially because the entanglement of formation is a measure that will be zero if and only if the state $\rho$ can be written as a mixture of product states, or in other words, if the system state  is separable. That is not true for the quantum discord, as separable states can have a non vanishing discord. In fact the only states for which the quantum discord $\delta^{\leftarrow}_{AB}$ is zero are states of the form \cite{vedra}
$\rho = \sum_{j}p_j\rho_j^{A}\otimes\ket{\psi_j}\bra{\psi_j}^{B}$.
Depending on the state the Eof can be significantly different from the discord, meaning that before the transformation from equation (\ref{eq:3}) to the equation (\ref{eq:5}) $\Delta$ could have been greater, less or equal to zero, since the balance was given as the difference between the entanglement of formation and the quantum discord of the system. 
\section{Structure of states saturating the b-SSA}
The structure of states that saturate the b-SSA can be obtained by using a theorem due to Petz \cite{petz}, regarding situations when the quantum relative entropy remains unchanged after the action of a certain map. That is possible because the conditional mutual information, as well as the other measures of quantum correlations in equation (\ref{eq:5}) can all be rephrased in terms of the quantum relative entropy.  To begin this analysis let us remind that the quantum discord is defined as the difference between the total correlations that a system share and the classical correlations in the system as in Eq. (\ref{qdisc}).
The mutual information is given by the relative entropy as \be I(A:B)=S(\rho_{AB}||\rho_{A} \otimes \rho_{B}),\ee while the classical correlation is given by Eq. (\ref{2}).
Using $\Phi_B(\rho_{AB}) = \sum_{i}p_{i}\rho_{A}^{i} \otimes \ket{\psi_{i}}_B\bra{\psi_{i}}$ and $\Phi_B(\rho_{B}) = \sum_{i}p_{i}\ket{\psi_{i}}_B\bra{\psi_{i}}$ we have that 
\be
J^{\leftarrow}_{AB}= S(\Phi(\rho_{AB})||\rho_{A} \otimes \Phi(\rho_{B})).
\ee
Therefore from (\ref{qdisc}) we have that
\br
\delta^{\leftarrow}_{AB} &=& \min_{\left\{ \Pi_{B}^i\right\}}\left[S(\rho_{AB}||\rho_{A} \otimes \rho_{B})\right.\nonumber \\&&- \left.S(\Phi_{B}(\rho_{AB})||\rho_{A} \otimes \Phi_{B}(\rho_{B}))\right].\label{dis1}
\er
In a similar fashion, for a tripartite system given by $\rho_{ABC}$,
\begin{eqnarray}
\delta^{\leftarrow}_{A(BC)} & = & \min_{\left\{ \Pi_{BC}^i\right\}}\left[S(\rho_{ABC}||\rho_{A} \otimes \rho_{BC})\right. \nonumber \\ 
&& - \left. S(\Phi_{BC}(\rho_{ABC})||\rho_{A} \otimes \Phi_{BC}(\rho_{BC}))\right],\label{dis2}
\end{eqnarray}
with $\Phi_{BC}(\rho_{ABC}) = \sum_{i}p_{i}\rho_{A}^{i} \otimes \ket{\psi_{i}}_{BC}\bra{\psi_{i}}$ and $\Phi_{BC}(\rho_{BC}) = \sum_{i}p_{i}\ket{\psi_{i}}_{BC}\bra{\psi_{i}}$.

For the Eof we have that 
\be\label{eq:11}
E_{AB} = \min_{\left\{ p_i,\ket{\psi_i}\right\}}\sum_ip_iS(\rho_{A}^i), 
\ee 
where the minimization is over all ensembles of pure states $\{p_i,\ket{\psi_i}\}$. Since $S(\rho_A|\rho_B) = - S(\rho_{AB}||\mathbf{1}_A\otimes\rho_B)$, we obtain
\be\label{eq:12}
E_{AB} = \min_{\left\{ \Pi_{B}^i\right\}}S(\Phi_B(\rho_{AB})||\mathbf{1}_A\otimes\Phi_B(\rho_B)),\ee
and through a similar derivation,
\be
\label{eq:13}
E_{A(BC)} = \min_{\left\{ \Pi_{BC}^i\right\}}S(\Phi_{BC}(\rho_{ABC})||\mathbf{1}_A\otimes\Phi_{BC}(\rho_{BC})).
\ee

By substituting Eqs. (\ref{dis1})--(\ref{eq:13}) in the b-SSA (\ref{eq:5}),
\be\label{eq:15}
I(A:C|B) \geq E_{AB} - E_{A(BC)} + \delta^{\leftarrow}_{A(BC)} - \delta^{\leftarrow}_{AB},
\ee
we can check that the mutual information, $I(A;C|B)$, is also present in the RHS, so it is easy to see that the condition for saturation of the b-SSA rests in the equality condition of the remaining terms of equation (\ref{eq:15}), i.e., necessarily   
\begin{multline}\label{16}
S(\Phi_{BC}(\rho_{ABC})||\rho_{A} \otimes \Phi_{BC}(\rho_{BC})) \\ = S(\Phi_{B}(\rho_{AB})||\rho_{A} \otimes \Phi_{B}(\rho_{B})),
\end{multline}
and
\begin{multline}\label{17}
S(\Phi_{BC}(\rho_{ABC})||\mathbf{1}_A\otimes\Phi_{BC}(\rho_{BC})) \\ = S(\Phi_B(\rho_{AB})||\mathbf{1}_A\otimes\Phi_B(\rho_B)).
\end{multline}

It is good to observe that in the relations written above we are not taking into account the minimizations that are taken on the POVM's. This is due to the fact that the same optimization is enacted in both terms for each equation so for a given optimum value the best basis is found and we can proceed with our analysis getting to the equations above. Continuing with our analysis, we know that the equality condition for the monotonicity of the relative entropy is guaranteed if there  exists a quantum operation $\hat{T}$ that maps $T\rho$ to $\rho$. So assuming that there exists a quantum operation that takes $B$ to $BC$ in the form of a recovery map $R_{B\rightarrow BC}$ \cite{petz}, we act over states $\Phi_{BC}(\rho_{ABC})$ and $\Phi_{B}(\rho_{AB})$ so that 
\be
R_{B \rightarrow BC}(\Phi_B(\rho_{AB})) = R_{B \rightarrow BC}\left(\sum_{i}p_{i}\rho_{A}^{i} \otimes \ket{\psi_{i}}_B\bra{\psi_{i}}\right),
\ee
and the structure of states that saturates the b-SSA will take the form
\be
\Phi_{BC}(\rho_{ABC})=\bigoplus_{j}q_{j}\sum_{i}p_{j|i}\rho_{A}^{i} \otimes \ket{\tilde{\psi_{j}}}_{b^L_{i}}\bra{\tilde{\psi_{j}}} \otimes \omega_{b^R_{i}C},
\ee
where $\omega_{b^R_{i}C} \in H_{b^R_{i}} \otimes H_{C}$, $\ket{\tilde{\psi_{j}}}_{b^L_{i}} \in H_{b^L_{i}}$ and $H_{B}= \bigoplus_{i} H_{b^L_{i}}\otimes H_{b^R_{i}}$. It is clear that by the possiblity of recovering the state belonging to $H_{ABC}$ from the state belonging to $H_{AB}$ makes a sufficient condition for us to call those kind of states as short Quantum Markov chains, similarly to the states 
that saturate the SSA \cite{hayden}. Equations (\ref{16}) and (\ref{17}) demand that $J^{\longleftarrow}_{A(BC)}=J^\leftarrow_{AB}$ which do not presents much relevancy, but it also demands that $E_{A(BC)}=E_{AB}$, that is, the entanglement of formation must be monogamous \cite{CKW} for those states. This can be very interesting in a quantum cryptographic setting were we are trying to minimize the access of third parties in a two part protocol.

\section{Quantum data processing}

Now we are going to see the implications of the b-SSA in a well know inequality, the quantum data processing inequality. The quantum data processing inequality was first introduced by Schumacher and Nielsen \cite{schuma}, where they also introduce a measure of entanglement - the Coherent Information, that obeys the data processing in the quantum regime. The coherent information is defined as \be I_c(A\rangle B) \equiv S(B) - S(AB),\label{ci}\ee i.e., the negative of the conditional entropy, $S(A|B)$ (which itself is negative when the system $AB$ is entangled). For the scheme in Fig. 1 the data processing is
\begin{equation}
I_{c}(A\rangle B_{1})\geq I_{c}(A\rangle B_{2}),\label{QDP}
\end{equation}
where there are two parties Alice and Bob and they share a bipartite state. Bob is the one that operates in his part of the state, in Fig. 1, and there are two stages corresponding to two operation in Bob's part. The first stage can be understood as the action of encoding information and produces $B_{1}$, the second stage could be some error correction to extract the information and yields $B_{2}$. Both environments start in a pure state and each interaction is unitary, guaranteeing the purity of the global state in all stages of the process. Also there is a change in the global state at those different stages - in the first the global system is $AB_{1}E_{1}$, and in the second part it is $AB_{2}E_{1}E_{2}$.
\begin{figure}
\begin{centering}
\includegraphics[scale=0.4]{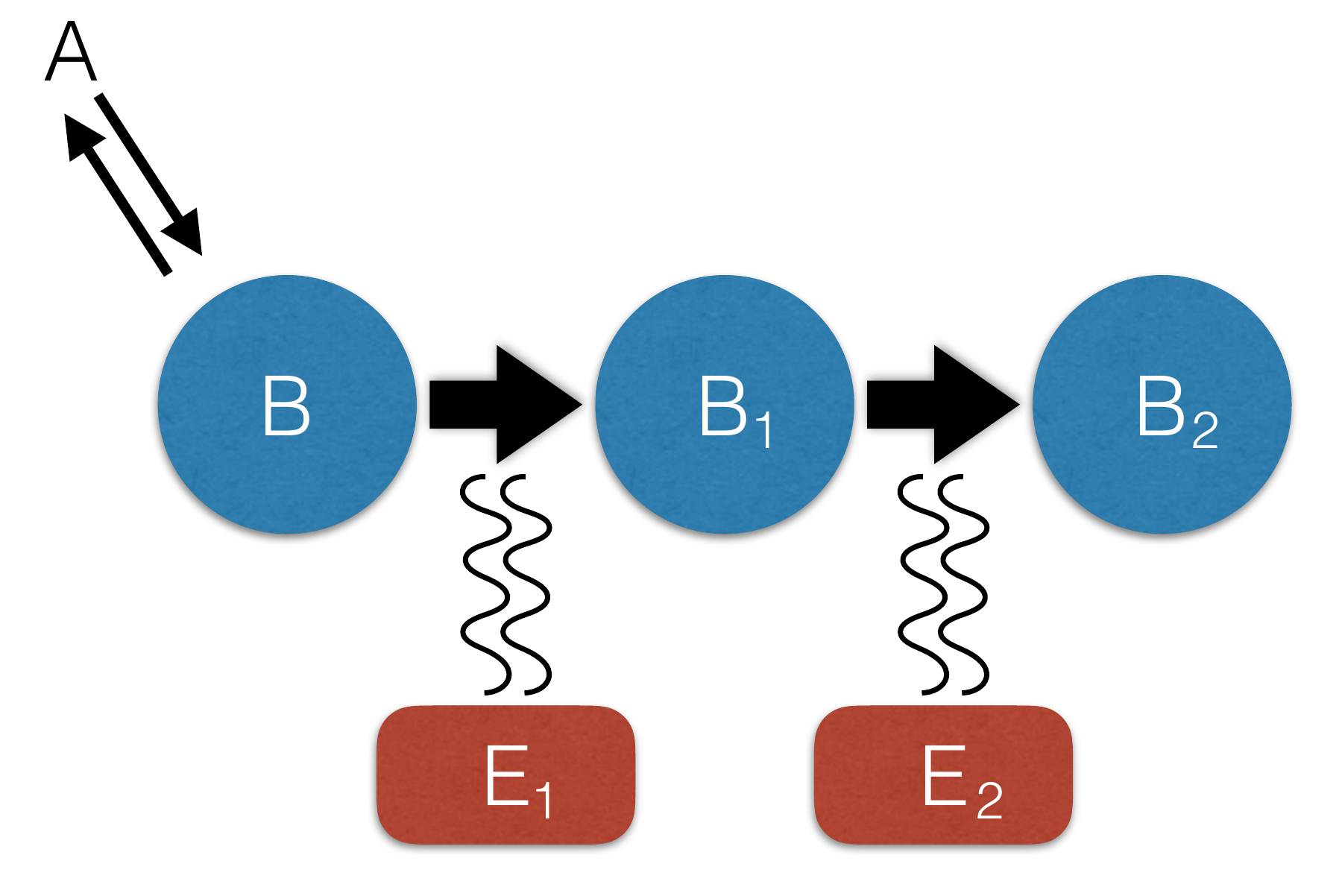}
\par\end{centering}
\caption{\footnotesize An illustration of a two stage noisy quantum channel. Alice ($A$) and Bob ($B$) share a bipartite state. Bob locally operates his state, which evolving as $B\rightarrow B_1\rightarrow B_2$ through the interaction with environments $E_1$ and $E_2$, respectively.} 
\end{figure}

The standard quantum data processing inequality (\ref{QDP}) says that in processing a quantum state we can only decrease quantum correlations between two parts, in agreement with its classical version. However, similarly to the SSA, the limiting bounds might change if quantum correlations are properly taken into account as we show now. 
By writing $I_{c}(A\rangle B_{1})$ and $I_{c}(A\rangle B_{1})$ explicitly in terms of the definition (\ref{ci}), and noticing that the quantum conditional mutual information, after the second process, is
\be\label{19}
I(A:E_{2}|E_{1})=S_{AE_{1}}+S_{E_{1}E_{2}}-S_{E_{1}}-S_{AE_{1}E_{2}},
\ee we see that 
\be\label{eq:10}
I_{c}(A\rangle B_{1})-I_{c}(A\rangle B_{2})=I(A:E_{2}|E_{1}).
\ee
By Eq. (\ref{eq:3}), and recalling that the global system ($AB_1E_1$, and $AB_2E_1E_2$) is pure, such that
$ E_{AE_1}-E_{A(E_1E_2)}+\delta^{\longleftarrow}_{A(E_1E_2)}-\delta^{\leftarrow}_{AE_1}=E_{AB_2}-E_{AB_1}-\delta^{\leftarrow}_{AB_2}+\delta^{\leftarrow}_{AB_1}
$, we can write \be\Delta=(E_{AB_{2}}-\delta^{\leftarrow}_{AB_{2}})-(E_{AB_{1}}-\delta^{\leftarrow}_{AB_{1}}),\ee in terms of $A$, $B_{1}$ and $B_{2}$ to obtain
\be
I_{c}(A\rangle B_{1})\geq I_{c}(A\rangle B_{2})+\Delta.\label{QDP2}
\ee
 This is the quantum data processing inequality when quantum correlations captured in $\Delta$ are appropriately included. Different situations of the inequality can be analyzed in terms of the quantum correlation shared between the subsystem $A$ with  $B_1$ and $B_2$.

We now analyze the balance of entanglement and QD in each of the stages between $A$ and $B_1$, and $A$ and $B_2$. We can  see that if the entanglement distributed in the system balances the quantum correlations (besides entanglement) we get the standard quantum data processing inequality. Otherwise the lower bound could be weaker or stronger. 
 Therefore if the Eof is equal to the QD in each stage, $\Delta = 0$,  and we recover the standard inequality. That happens when both the systems $AB_1$ and $AB_2$ are pure, possible only when the environments $E_1$ and $E_2$ are uncorrelated (not even classically) from $B_1$ and $B_2$, respectively. Therefore the evolution $AB\rightarrow AB_1\rightarrow AB_2$ is unitary.  Of course this is not the only situation where $\Delta=0$ - it might happen that the exceeding correlation (entanglement) in one stage cancels out the exceeding correlation (entanglement) at the other stage. There are many situations when this is possible whenever the state of $AB_1$ and $AB_2$ alone are mixed. Therefore we assume that the standard result of the quantum data processing applies specifically in these cases. 
 The situation when $\Delta > 0$ is extremely interesting as it imposes a stronger lower bound to the data processing inequality. Rephrasing its meaning, the quantum data processing inequality (\ref{QDP2})
 says that in processing a quantum state we can only decrease the quantum  correlations  between  two  parts, and the amount of this decreasing is bounded by the balance of quantum correlations in the process. In contrast, if
 $\Delta < 0$, in principle the processing could be improved. However, since this lower bound is weaker than the standard quantum data processing, it is not a relevant case, but it means that the correlations at the final stage could be larger than initially, as if the environments were contributing to the processing with an extra amount of quantum correlations making the processing better.
This last case can be considered non-physical, since there are different proofs attesting the non-negativity of the quantum data processing inequality, and the bound being less than zero would violate the standard inequality. 

Lastly, a intuition can be given by a different lower bound for the data processing in terms of the flow of locally inaccessible information\cite{fan}, as in Fig. 2. Noting that the difference between the Eof and the QD can be written as 
\begin{equation}
E_{AB_{2}}-\delta^{\leftarrow}_{AB_{2}}\equiv \frac{1}{2}(\mathcal{L}_{E_{1}E_{2}\rightarrow A\rightarrow B}-\mathcal{L}_{B\rightarrow A\rightarrow E_{1}E_{2}}).\label{LII}
\end{equation}
The relation on the right hand side of (\ref{LII}) represents the net flow of locally inaccessible information (LII)
\begin{equation}
\mathcal{L}_{R\{E_{1}E_{2}\}}\equiv \frac{1}{2}(\mathcal{L}_{E_{1}E_{2}\rightarrow A\rightarrow B}-\mathcal{L}_{B\rightarrow A\rightarrow E_{1}E_{2}}), 
\end{equation}
from $\{E_{1}E_{2}\}$ to $A$ to $B$ and from $B$ to $A$ to $\{E_{1}E_{2}\}$.
\begin{figure}
\begin{centering}
 \includegraphics[scale=0.4]{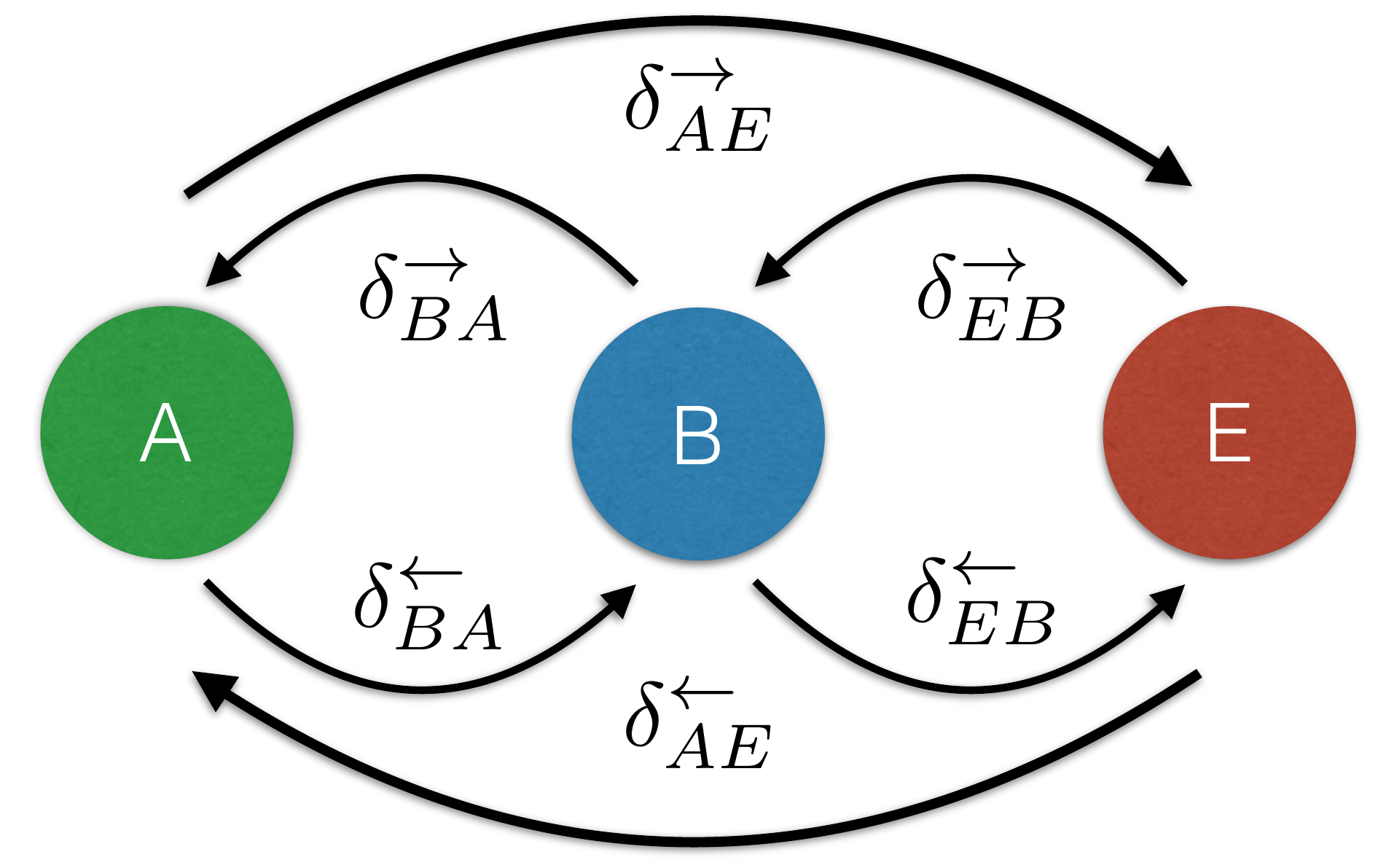}
\par\end{centering}
\caption{\footnotesize A schematic representation of the cycle of the Locally Inaccessible Information (LII), measured by Quantum Discord.} 
\end{figure}
The notation $R\{E_{1}E_{2}\}$ is there to specify that the net flow of LII is in respect with both environments in each stage of the processing, while $R\{E_{1}\}$ implies a net flow in respect with the environment in the first stage only. By Eq. (\ref{eq:15}) we establish a lower bound for the quantum data processing inequality based on the difference of the net flow of LII in an out of the environments $E_{1}$ and $E_{2}$ as
\be
I_{c}(A\rangle B_{1})-I_{c}(A\rangle B_{2})\geq\mathcal{L}_{R\{E_{1}E_{2}\}}-\mathcal{L}_{R\{E_{1}\}}.
\ee
The flow of LII is based in measurements of the quantum discord viewed in tripartite systems, where those measurements are taken from bipartite sides in both directions capturing only the quantum correlations. By those contributions while most of the exchange of LII is happening from $B$ to the environments some exchange is happening from the system $A$ and the environments, even though there is no operation in $A's$ part of the state. 

The intuition 
is that the locally inaccessible information in respect to the whole processing $\mathcal{L}_{R\{E_{1}E_{2}\}}$ should be greater than the locally inaccessible information on the first stage $\mathcal{L}_{R\{E_{1}\}}$ since it should be harder to disturb the system after being processed twice, and the bound should be greater than zero. But noticing what the standard quantum data processing inequality tells, we believe that while the bound is greater than zero the locally inaccessible information acquired after both stages is not usable nor by Alice or by Bob, since it is not locally accessible. This is different than saying that the LII is destroyed during the processing, because it is only not accessible by both parts. The question then would be if there is a way to harness the extra LII in order to strengthen the standard inequality.

\section{conclusion} 

Starting from the bounded weak monotonicity we derived an equivalent inequality that we called bounded strong subadditivity (SSA) of the von Neumann entropy. We showed that the lower bound obtained for the SSA can take a range of values, positive, negative or null. Depending on the mixture of states utilized it can give a stronger bound than usual to the quantum conditional information, since the lower bound is written as a balance of quantum correlations on the system described by the entanglement of formation and quantum discord. Both measures, the Eof and the QD were rewritten as relative entropies in order to use Petz's theorem \cite{petz} and the Hayden et al. \cite{hayden} result to obtain the the structure of states that would saturate the bounded quantum conditional entropy. The resulting states exhibit the form of short quantum Markov chains similarly to the states that saturate the standard strong subadditivity, in the aspect that we can recover the global state from a reduced form. This structure also demanded the entanglement of formation to respect a monogamous relation, which can make those states useful in cryptography protocols. In addition we examined the consequences of the bounded SSA in the quantum data processing inequality, although it is not clear why the data processing should not be stricter than usual given the bound in terms of the Eof and the QD, a lower bound in terms of the difference in net flow of locally inaccessible information was achieved given additional insight. Even though the bound is greater than zero, it is possible that the locally inaccessible information seen by Bob is not extractable or useful in the processing, being only possible to use the locally accessible part. The question of weather it is conceivable a protocol where we can use the LII in the system and the quantum data processing is violated remains for further investigation.

\section*{Acknoledgements}{This  work  is  partially  supported  by  the  Brazilian  National  Council  for Scientific and Technological Development (CNPq),  Coordination for the Improvement of Higher Education Personnel (CAPES) and FAPESP through the Research Center in Optics and Photonics (CePOF).}

\bibliographystyle{apsrev4-1} 

\end{document}